\documentstyle[12pt]{article}
\oddsidemargin--1mm
\begin{document}
\newcommand{\pl}{\partial}
\newcommand{\be}{\begin{equation}}
\newcommand{\ee}{\end{equation}}
\newcommand{\ba}{\begin{eqnarray}}
\newcommand{\ea}{\end{eqnarray}}
\newcommand{\mbf}[1]{\mbox{\boldmath$ #1$}}
\def\<{\langle}
\def\>{\rangle}

\noindent
\hspace*{10cm} {\small\sf IFT preprint UFIFT-MATH-01-15}

\begin{center}
{\large\bf Infared Yang-Mills theory as a spin system.

\vskip 0.2cm
 A lattice approach }

\vskip 0.5cm
Sergei V. Shabanov {\footnote{on leave from Laboratory of
Theoretical Physics, JINR, Dubna, Russia}}

\vskip 0.2cm

{\em Institute for Fundamental Theory, 
Departments of Physics and Mathematics,\\ University of Florida,
Gainesville, FL- 32611, USA}
\end{center}

\begin{abstract}
To verify the conjecture that Yang-Mills theory in the infrared limit
is equivalent to a spin system whose excitations are knot solitons,
a numerical algorithm based on the inverse Monte Carlo method is proposed.
To investigate the stability of the effective spin field action, 
numerical studies of the renormalization group flow for the coupling 
constants are suggested. A universality of the effective
spin field action is also discussed.   

\end{abstract}

{\bf 1}. It was conjectured \cite{fn1}
that the SU(2) Yang-Mills theory in the 
infrared region can be described as a spin system with
the following action
\be
S = \int dx\left\{
m^2(\pl_\mu\mbf{n})^2 + g^{-2}[\mbf{n}\cdot(\pl_\mu\mbf{n}
\times\pl_\nu\mbf{n})]^2 
\right\}   \ ,
\label{1}
\ee
where $\mbf{n}^2=1$ (boldface letters stand for three-vectors). 
The path integral representation 
of the effective action (\ref{1}) 
can be deduced from the Yang-Mills theory path integral 
via an implicit change of integration variables \cite{svs1}. 
The analysis can be extended to the SU(N) case \cite{svs2,fn2}.
Nonperturbative excitations of the effective theory
are knot solitons \cite{knots}. Knot solitons look more like stringy
excitations, which is believed to be a right physical picture of
nonperturbative excitations of gauge fields. Yet, if the effective action
(\ref{1}) turns out to be a good approximation to the Yang-Mills theory 
in the infrared limit, the nonperturbative dynamics can be studied 
by quantum soliton theory methods. The mass gap in the spectrum of 
quantum Yang-Mills theory would therefore naturally be introduced 
as the lowest energy bound in the quantum soliton spectrum. A few important
questions are to be addressed to validate or invalidate this attractive 
picture. First, what is the actual value of the mass scale $m^2$ which
determines the low energy bound (in the classical soliton theory) \cite{bound}?
Second, is the effective action stable from the point of view of the 
renormalization group flow of its coupling constants? Third, 
how big are the higher order corrections to (\ref{1})? 
The purpose of this paper is to set up a numerical
approach to answer these questions.     

The existence of degrees of freedom whose dynamics dominates in
the infrared region of Yang-Mills theory was established 
in numerical simulations \cite{monlat,pol,mp}
of lattice Yang-Mills theories some time ago. 
It was observed 
that dominant contributions to the  string tension
come from topological defects (monopoles)  
which occur in typical vacuum configurations of gauge fields
when the latter are taken in a special gauge known as the maximal
Abelian gauge. Topological defects unavoidably occur in any gauge that breaks
the gauge group to its maximal Abelian subgroup \cite{thooft}.
In classical theory, it is evident from the fact that the homotopies
of the gauge group and its maximal Abelian subgroup are different \cite{singer}
(see more on the gauge fixing problem in quantum gauge theories in \cite{pr}).
The importance of the above numerical discovery is 
that the defects alone are sufficient to reproduce 
essential nonperturbative features of Yang-Mills theory.
The numerical procedure of singling out topological defects in Abelian
projections of the lattice Yang-Mills theory  can therefore be used 
in a theoretical analysis to parameterize the 
relevant degrees of freedom of the Yang-Mills connection 
and compute their effective action. 
It should be noted that the gauge fixing here does not serve its 
conventional purpose -- removing nonphysical degrees of freedom --
but rather it becomes an auxiliary tool to identify the  
degrees of freedom relevant for the infrared physics
of the Yang-Mills theory.
After a change of variables, that splits all the Yang-Mills degrees of freedom
into the ``infrared relevant'' ones and the rest, is found, the effective
action can be computed in any convenient gauge and its gauge
invariance can be established by the standard BRST technique \cite{svs1,svs2}.  

In our earlier works \cite{svs1,svs2},
a relation between the spin field $\mbf{n}$ and  topological
defects of the connection in the maximal Abelian gauge has been found
\be
\mbf{A}_\mu = 
g^{-1} \pl_\mu \mbf{n}\times\mbf{n} + \mbf{n}C_\mu +\mbf{W}_\mu\ ,
\label{2}
\ee
where $\mbf{W}_\mu$ satisfies 
the following conditions.
It is perpendicular to $\mbf{n}$ and
\be
\pl_\mu\mbf{W}_\mu + (\mbf{\alpha}_\mu +\mbf{n}C_\mu)\times\mbf{W}_\mu \equiv
\mbf{\nabla}_\mu(\mbf{\alpha} +\mbf{n}C) \mbf{W}_\mu= 0\ .
\label{3}
\ee
Here $\mbf{\alpha}_\mu = g^{-1} \pl_\mu\mbf{n}\times\mbf{n}$ is the 
connection introduced in \cite{cho}. 
Relation (\ref{2}) is a change of variables in the space of connections.
Indeed, the original variables $\mbf{A}_\mu$ have 12 independent scalar
functions. There are two independent scalar functions in $\mbf{n}$, four
in $C_\mu$, and, hence, there must be six independent scalar functions in
$\mbf{W}_\mu$. This is the case indeed 
because 12 components of $\mbf{W}_\mu$
satisfy six independent conditions: Four in $\mbf{n}\cdot\mbf{W}_\mu=0$ and
two in (\ref{3}). Note that if $\mbf{W}_\mu$ is perpendicular to $\mbf{n}$,
then covariant derivatives of $\mbf{W}_\mu$ with respect the 
connection $\mbf{\alpha}_\mu +\mbf{n}C_\mu$ is always perpendicular 
to $\mbf{n}$. The inverse transformation can be found by multiplying
(\ref{2}) by $\mbf{n}$ using first the dot and then cross products.
The obtained relations allow one to express $C_\mu$ and $\mbf{W}_\mu$
as functions of $\mbf{A}_\mu$ and $\mbf{n}$. Substituting them
into (\ref{3}), an equation for $\mbf{n}$ as a functional of $\mbf{A}_\mu$
is derived.  
It has been shown that for a given $\mbf{A}_\mu$, the corresponding
spin field can be computed as \cite{svs1}
\be
\mbf{n} =\textstyle{\frac 12} 
{\rm tr}\, (\mbf{\tau}\Omega_{A}^\dagger\tau_3\Omega_A)\ ,
\ \ \ \ \mbf{\tau} = (\tau_1, \tau_2, \tau_3)\ ,
\label{4}
\ee
where $\tau_i$ are the Pauli matrices, ${\rm tr\,}(\tau_i\tau_j)=
2\delta_{ij}$ and $\Omega_A$ is a group element which depends on
$\mbf{A}_\mu$ so that the gauge transform
of $\mbf{A}_\mu$ with $\Omega_A$ satisfies the maximal Abelian
gauge. Topological numbers of the defects
have an integral representation via the spin field $\mbf{n}$ 
\cite{svs1}.

The gauge transformation $\Omega_A$ is in general
singular (it might not even be single valued
in spacetime). In other words, 
for a typical vacuum configuration of gauge fields,
the maximal Abelian gauge can {\it only} be achieved by a singular
gauge transformation.
According to numerical simulations \cite{monlat}, the third 
component of the gauge fixed configuration 
$\mbf{A}_\mu^\Omega$, which is associated with the unbroken $U(1)$
subgroup, carries Dirac magnetic monopoles. The monopoles alone contribute about
90\% to the energy of the flux tube (string) between static sources  
in the Yang-Mills theory. The monopole spacetime
trajectories  are determined by singularities of $\Omega_A$. Hence, by taking
Abelian connections of the monopoles and applying the inverse gauge 
transformation $\Omega_A^\dagger$ to them, one can parameterize the relevant
(or ``monopole producing'') degrees of 
freedom of a generic Yang-Mills connection.
The result of this procedure is given by Eqs. (\ref{2}) and (\ref{3}). 
By construction, the spin field $\mbf{n}$ carries all the information
about the spacetime distribution  of the defects, and, hence,
its effective theory should describe the infrared
physics of the original Yang-Mills theory. 

Using the path integral representation of the spin field effective 
action \cite{svs1}, one-loop calculations have been done in \cite{cern}
\footnote{see also \cite{ln} where  different approaches, not
related to the observations in lattice gauge theories, have been explored.}.
They show that the action (\ref{1}) is to be modified by adding 
terms containing higher order time derivatives (which might be a
source of instability of knot solitons). The one-loop renormalization
group flow for the effective action parameters also indicates that
the effective action might be stable in the infrared region \cite{cern}. 
The results of \cite{cern} seems encouraging and deserve 
further studies, particularly, by some nonperturbative methods.

Since the topological defects in the gauge-fixed theory are not solutions
to the classical equations motion with a finite energy or action, 
it is rather hard to give them an accurate mathematical meaning in 
continuum quantum theory.
The nonlinear change of variables (\ref{2})  makes sense
for classical connections whose values are well defined almost everywhere
in spacetime. 
Typical quantum fields that form support of the path integral
measure are distributions rather than classical smooth functions. 
The change of variables (\ref{2}) is, in fact, 
ill-defined (because it involves products of distributions) 
unless some short-distance regularization is implemented.
This can be achieved either by defining the path integral
perturbatively with an ultraviolet cutoff, or by using the
lattice (nonperturbatively  defined) path integral for gauge theories.
This latter approach is adopted in the paper.  
 
Here we develop the idea, first suggested  in \cite{svs1}, 
of using the inverse Monte Carlo method \cite{sw} 
to find out whether  (\ref{1}) is indeed a good approximation to the 
infrared Yang-Mills theory.
Within the framework of lattice Yang-Mills theory, 
an explicit numerical algorithm is proposed 
to compute and study the effective action for the spin field and
the renormalization group flow of its parameters, which comprises the main
goal of the paper.

{\bf 2}. In lattice gauge theories, the dynamical variables 
are group elements $u_{x\mu}\equiv u_l\in SU(2)$ associated with each link,
where $x$ enumerates lattice sites, and $\mu$ indicates the direction
of the link from the site $x$.  Let $\{u_l\}$ be a Wilson ensemble of
link variables distributed with the Boltzman
probability $Z_W^{-1}e^{-S_W}$ with $S_W$ being the Wilson action of 
link variables and $Z_W$ the normalization factor (the partition function). 
The first step of the proposed numerical simulations
is to generate an ensemble
of the spin field from the Wilson ensemble.
The spin field in the decomposition (\ref{2}) is defined
by Eq. (\ref{4}). Thus, for every configuration $u_l$ one has to find
a configuration of gauge group elements $\Omega_x(u_l)$ such that
the gauge transformed configuration 
\be 
u_l^\Omega = \Omega_x u_l \Omega_{x+\mu}^\dagger
\label{gt}
\ee
satisfies the maximal Abelian gauge. Here $x+\mu$ denotes the lattice
site next to $x$ in the direction $\mu$.
The group elements $\Omega_x(u_l)$
can be found by maximizing the function \cite{wiese}
\be
\chi_u(\Omega) = \textstyle{\sum_l}{\rm tr }\,\left[
\tau_3\left(u_l^{\Omega}\right)^\dagger\tau_3 u_l^\Omega\right]
\label{chi}
\ee
for each configuration $u_l$ of the Wilson ensemble. 
The collection of group elements $\Omega_x$ at all lattice sites is 
regarded as variables, while $u_l$ are just parameters. 
For every configuration $u_l$ the function $\chi_u$ can have
many local maxima. This is an evidence
of the Gribov problem in lattice gauge theories (see for a review \cite{pr}
and references therein).   
For every element of the Wilson ensemble $u_l$, one should take $\Omega_x(u)$
at which $\chi_u$ attains its absolute maximum. Finding an absolute
maximum of $\chi_u$ is a difficult, if not impossible, task 
in the numerical gauge fixing. The state-of-the-art extrapolation
toward the global maximum of $\chi_u$ can be found in \cite{mp}. 
The ensemble of
the spin field is then computed as  
$\mbf{n}_x(u_l) =\frac 12{\rm tr}\,(\mbf{\tau}\Omega^\dagger_x
\tau_3\Omega_x)$. 

It is also possible to find the lattice version of the change of variables
(\ref{2}) and therefore to obtain a system of cubic 
equations whose solution
defines the spin field components as functions of link variables
(the lattice analog of the equations for the spin field
suggested in \cite{svs1}). 
Define 
an algebra element at each site $n_x =\Omega_x^\dagger\tau_3\Omega_x$
which satisfies the constraint ${\rm tr}\,n_x^2 =2$. Combining (\ref{gt})
and (\ref{chi}) and introducing the Lagrange multiplier $\xi_x$ to
take into account the constraint on $n_x$, the extreme value problem
for (\ref{chi}) is equivalent to the extreme value problem for
the function
\be
\tilde{\chi}_u(n) = \textstyle{\sum_x}\left\{ 
\textstyle{\sum_\mu}{\rm tr }\,\left[
n_{x+\mu}u_l^\dagger n_x  u_l\right] + \xi_x \left(
\textstyle{\frac 12} {\rm tr }\,n_x^2 -1
\right)\right\}\ .
\ee 
Setting the variations of $\tilde{\chi}_u(n)$ with respect to $n_x$ and
$\xi_x$ to zero, the following equation for $n_x$ can be deduced
\be
\varphi_x(n,u)\equiv
\textstyle{\sum_\mu}\left(
 u_{x-\mu,\mu}^\dagger n_{x-\mu} 
u_{x-\mu,\mu} + u_{x,\mu} n_{x+\mu}u_{x,\mu}^\dagger
\right) + 2\xi_xn_x = 0\ . 
\label{nfield}
\ee
Now assume that $\Omega_x(u)$
is a local maximum of (\ref{chi}) and, hence, $n_x$ is a solution to
(\ref{nfield}), then $u_l^\Omega$ satisfies the maximal
Abelian gauge and, by construction, $n_x = \mbf{n}_x\cdot \mbf{\tau}$. 
The Lagrange multiplier $\xi_x=\xi_x(u,n)$
is fixed by multiplying (\ref{nfield}) by $n_x$
and taking the trace.
After the substitution of $\xi_x =\xi_x(u,n)$ into (\ref{nfield}), one
gets the equation for the components of the spin
field. Only two scalar equations in (\ref{nfield})  are independent. 
They determine two independent components of the spin field as functions 
of the link variables.
 
As has been mentioned above, 
the group elements $\Omega_x$ and hence the spin field
are not regular everywhere in space 
in the continuum theory. 
It is not difficult to give an example of $\Omega_x$ and the 
corresponding spin field $\mbf{n}_x$ such that the connection 
$\mbf{\alpha}_\mu$ coincides with the Wu-Yang monopole (e.g.,
take $\mbf{n}=\mbf{x}/r, \ r=|\mbf{x}|$) which, in the maximal Abelian gauge,
contains a Dirac monopole at the origin. The spin field
is ill-defined at the position of the monopole.
In the lattice gauge theory, the topological defects occur on the dual
lattice sites \cite{wiese}. 
For every configuration $u_l$, the configuration $\Omega_{x}(u_l)$ and, hence,
$\mbf{n}_x$ are well defined and contain 
all the information about locations of the defects
(magnetic monopoles)
on the dual lattice  and their topological numbers (magnetic 
charges). Consider an elementary cube of the spatial lattice and 
spins on its vertices. Frankly speaking, 
with an isolated defect present at the cube center, 
the spins are directed outward (or inward) the cube. 
In the confinement phase the monopole-antimonopole  pairs 
(or monopole loops, when describing topological defects by
their spacetime trajectories) are condensed (no isolated defects), 
therefore the above simple
visual picture would not be valid. However, $\mbf{n}_x$ is still 
well defined at each lattice site and its dynamics can be studied.

{\bf 3}. The configurations of gauge fields $u_l$ are
distributed with the Boltzman probability $Z_W^{-1}e^{-S_W}$. 
The spin field configurations obtained
from the Wilson ensemble must also be distributed with some probability
$Z_s^{-1}e^{-S}$ where $S$ is the unknown effective action of the spin field. 
The problem is therefore: Given
an ensemble of $\mbf{n}_x$, find the corresponding probability or
the effective action $S(\mbf{n})$.

Any correlator of the spin field can be computed  
by the Monte Carlo method since the ensemble is known
\be
\< F(\mbf{n})\>_n \equiv Z^{-1}_s\int D\mbf{n} e^{-S} F(\mbf{n}) =
\frac{1}{M}\sum_{\{\mbf{n}\}} F(\mbf{n}) + O(M^{-1/2})\ ,
\label{fav}
\ee
where the sum is taken over the ensemble of the spin field,   
$Z_s=\int D\mbf{n} e^{-S}$ 
is the partition function, $D\mbf{n}=\prod_xd\mbf{n}_x$
and the integration over a spin at each site implies the integration
over a unit two-sphere. Parameterizing the spin vector by the spherical
angles
\be
\mbf{n}_x=(\cos\phi_x\sin\theta_x,\, \sin\phi_x \sin\theta_x,\, \cos\phi_x)
\label{sphere}
\ee
we get $d\mbf{n}_x=d\phi_xd\theta_x\sin\theta_x$ where $\phi_x\in [0,2\pi)$ and
$\theta_x\in [0, \pi]$. 
The expectation value (\ref{fav}) is also realized as
an expectation value with respect to the original Wilson ensemble. Note that
Eq.(\ref{nfield}) defines the spin field as a function of
link variables $\mbf{n}_x=\mbf{n}_x(u)$. Hence,
\be
\< F(\mbf{n})\>_u = Z^{-1}_W \int Du e^{-S_W(u)}F(\mbf{n}(u))\ ,
\label{ym}
\ee
where $Z_W$ is the partition function for the Wilson action.     
In principle, this observation
can be used to determine 
the effective action directly via the original Wilson
ensemble. The idea is the same as in the continuum case \cite{svs1}.
Define the function $\Delta(u,\mbf{n})$ by the condition 
\be
\int D\mbf{n}\, \Delta(u,\mbf{n})
\prod_x\delta(\mbf{\varphi}_x(\mbf{n},u)) =1\ , 
\label{iden}
\ee
where $\mbf{\varphi}_x = \frac 12{\rm tr}(\mbf{\tau}\varphi_x)$ (see
(\ref{nfield})), which leads
to  $\Delta(u,\mbf{n}) =\det (\pl \varphi_x^a/\pl n_y^b)$.
The parameterization (\ref{sphere}) must be used to compute the derivatives
and also the identity $\mbf{\varphi}_x\cdot\mbf{n}_x =0$ is to be taken
into account when computing the determinant. In spacetime the matrix
 $\pl \varphi_x^a/\pl n_y^b$ appears to be sparse because nonzero
elements can only occur for $y=x, x\pm\mu$. 
Substituting the identity
(\ref{iden}) into the integrand in (\ref{ym}), changing the order of integration
and comparing it with (\ref{fav}), it is not difficult to deduce that
\be
S(\mbf{n}) = -\ln \left\< \Delta(u,\mbf{n})
\prod_x\delta(\mbf{\varphi}_x(\mbf{n},u))\right\>_u\ ,
\label{exacts}
\ee     
where $\mbf{n}_x$ is now held fixed in the average over the Wilson
ensemble. 

Although (\ref{exacts}) defines the spin field effective action 
as an expectation value of some function of the link variables,
we are interested only in its behavior in the infrared region.
In the continuum case this amounts to the so called gradient 
expansion of the (nonlocal) effective action.
In the numerical approach, the action is sought in the form
\be
S \approx \sum_i\, \lambda_iS_i
\ee
where $S_i$ are some specified functions of the spin field, whereas the 
coupling constants $\lambda_i$ are to be determined.  Since we are interested
to compare the effective action with (\ref{1}), we first find all possible
local and independent terms that are of the forth order in derivatives
and might contribute to the gradient
expansion of $S$. In addition to the two terms in (\ref{1}) it is necessary
to include $\pl_\mu^2\mbf{n}\cdot\pl_\nu^2\mbf{n}$ and
$[(\pl_\mu\mbf{n})^2]^2$. The four terms
are all independent Lorentz and isotopic invariant terms containing
up to four derivative operators. In fact, there is one more 
invariant term which can be built by contracting the Lorentz tensor
$\pl_\mu\mbf{n}\times\pl_\nu\mbf{n}$ with its dual (like a $\theta$-term
in the Yang-Mills action). Higher order terms can be classified
accordingly by contracting invariant irreducible tensors of the isotopic and
Lorentz  groups with isotopic tensor products of the spin field and its
derivatives. Then the renormalization group flow of 
the constants $\lambda_i$ must be 
studied as high momentum components of the spin field are removed
(integrated out). The renormalization
group flow would show whether or not the effective action is stable
(or, in other words, is a good approximation to (\ref{exacts})) 
in the infrared limit, and thereby validate or invalidate the conjecture.  

We set
\be
S_i = \frac{1}{m_i}\sum_x S_{ix}
\label{action}
\ee
where $m_i $ is the number of spins involved into a local interaction $S_{ix}$.
Let $\mbf{s}_{x,\mu}= 
\mbf{n}_{x+\mu}-\gamma_{x\mu}\mbf{n}_{x}$, where $\gamma_{x\mu}=
\mbf{n}_{x+\mu}\cdot\mbf{n}_x$ is defined
by the condition $\mbf{s}_{x,\mu}\cdot\mbf{n}_x =0$ to make the correspondence
with the continuum theory $\pl_\mu\mbf{n}\cdot\mbf{n}=0$. Let $a$ be a
lattice spacing.
The local spin interactions are written as 
\ba
S_{1x} &=& a^2\sum_\mu\left[\mbf{s}_{x,\mu}^2+
 \mbf{s}_{x-\mu,\mu}^2\right]\ ;
 \label{s1}\\
S_{2x}&=& \left[\sum_\mu \mbf{s}_{x,\mu}^2\right]^2 +
\left[\sum_\mu \mbf{s}_{x-\mu,\mu}^2\right]^2\ ;\\
S_{3x}&=& \sum_{\mu,\nu}\left[ (\mbf{s}_{x,\mu}
\times \mbf{s}_{x,\nu})^2 +
(\mbf{s}_{x-\nu,\mu}\times\mbf{s}_{x-\nu,\nu})^2 +
(\mbf{s}_{x-\mu,\mu}\times\mbf{s}_{x-\mu,\nu})^2
\right]\ ;\\
S_{4x}&=& \left[\sum_{\mu}(\mbf{s}_{x,\mu} -\mbf{s}_{x-\mu,\mu})\right]^2+
\left[x\rightarrow x^\prime = x+\mu
\right] + 
\left[x\rightarrow x^\prime = x+\mu\right] \ . \label{s4}
\ea
In the continuum limit, the action (\ref{action}) goes into (\ref{1})
with the additional terms described above. The $\theta$-term has 
the same form as $S_{3x}$ where instead of the sum of squares,
the sum of the dot products of
each vector and its Lorentz dual has to be taken.
Local spin interactions giving rise to terms with higher powers of
$\pl_\mu$ in the continuum limit can be constructed similarly by
using the correspondence rule: $\pl_\mu\mbf{n}\rightarrow\mbf{s}_{x,\mu}$,
$2\pl_\mu\pl_\nu\mbf{n}\rightarrow
\mbf{s}_{x+\mu,\nu}-\mbf{s}_{x,\nu} +  \mbf{s}_{x+\nu,\mu}-\mbf{s}_{x,\mu}$,
etc. 

Note that two terms in $S_{1x}$
give the same contribution as the sum over $x$ is taken in 
(\ref{action}) and therefore $m_1 =2$. Similarly,
each of three terms in $S_{ix}$ $(i=2,3,4)$ gives the same contribution
to   (\ref{action}) and $m_i =3$. The reason the equivalent term are
given in $S_{ix}$ is that 
$S_{ix}$ is the part of the action $S_i$ that
contains all terms involving the spin $\mbf{n}_x$ at a fixed site $x$.      
This representation will be useful in what follows.

{\bf 4}. Here we formulate the inverse Monte Carlo algorithm for
computing $\lambda_i$. The inverse Monte Carlo method is well known 
in studies of the real space renormalization group of spin systems \cite{sw}.
It has also been applied to compute an effective action for monopole
currents in the maximal Abelian projection \cite{monte}.
The use of the spin field order parameter $\mbf{n}_x$ rather than the monopole
current is more appealing because of several reasons (relations to the quantum
soliton theory, similarities between strings and knot solitons) pointed 
out after (\ref{1}).

Let $S_x$ denote all terms in $S$ that contain the spin $\mbf{n}_x$
at a fixed site $x$, $S_x=\sum_i \lambda_iS_{ix}$. For every $S_i(\mbf{n})$
we construct a new function 
\be
\bar{S}_i(\mbf{n},\lambda)= \frac{1}{m_i} \sum_x Z_x^{-1}\int d\mbf{n}_x
e^{-S_x} S_{ix} \equiv \sum_x \bar{S}_{ix}
\label{id1}
\ee
where $Z_x=Z_x(\mbf{n}) =\int d\mbf{n}_x e^{-S_x}$. 
The bar in  $\bar{S}_{ix}$ denotes an expectation value carried out 
with respect to the effective action $\<\cdots\>_n$ but calculated 
for only one spin, $\mbf{n}_x$. The environment (i.e. neighboring spins)
is held fixed. So, $\bar{S}_{ix}$
depends only on the spins at the sites neighboring with $x$, i.e.,
on $\mbf{n}_{x\pm\mu}$ and $\mbf{n}_{x\pm 2\mu}$. Taking the expectation
value of $\bar{S}_i$, we find the identity
\be
\<S_i(\mbf{n})\>_u = \<\bar{S}_i(\mbf{n},\lambda)\>_u\ .
\label{id2}
\ee     
Using the Monte Carlo method (\ref{fav}), 
the l.h.s. of (\ref{id2}) can be computed,
while the r.h.s. cannot. The integral over $\mbf{n}_x$ 
in (\ref{id1}) cannot be computed for given configurations 
of neighboring spins because the true values of $\lambda_i$ are not known.
Had the coupling constants been known, the ordinary integral in (\ref{id1})
could have been computed, for instance, numerically for any given 
$\mbf{n}_{x\pm \mu}$ and $\mbf{n}_{x\pm 2\mu}$. 

Suppose some trial values $\tilde{\lambda}_i$ of the coupling
constants are taken to compute the r.h.s. of (\ref{id1}). The equality
\be
\<\bar{S}_i(\mbf{n},\lambda)\>_u =  \<\bar{S}_i(\mbf{n},\tilde{\lambda})\>_u
\label{id3}
\ee
holds if and only if $\lambda_i=\tilde{\lambda}_i$. 
This is used to set up
an iterative algorithm to find the true coupling constants. 
Equation (\ref{id3}) is regarded as a system of nonlinear equations
where the l.h.s. is known (cf. (\ref{id2})). It can be solved 
numerically by Newton's method or some of its alterations.
For $\tilde{\lambda}_i\approx\lambda_i$ we have
\be
\<S_i\>_u -\< \bar{S}_i(\mbf{n},\tilde{\lambda})\>_u \approx\sum_j
\left.\left\<\frac{\pl}{\pl\lambda_j}\, \bar{S}_i(\mbf{n},{\lambda})\right\>_u
\right\vert_{\lambda=\tilde{\lambda}} (\lambda_j -\tilde{\lambda}_j)  
\ee
Using the definition (\ref{id1}), it is not difficult to show
that 
\be
  \frac{\pl}{\pl\lambda_j}\, \bar{S}_i = \bar{S}_j\bar{S}_i -
\overline{S_jS_i} \ .
\ee 
The function  $ \overline{S_jS_i}$ is 
defined by (\ref{id1}) where $S_{ix}$
is replaced by $S_{jx}S_{ix}$.   
The true values of the coupling constants are computed by the 
iterating procedure
\ba
b_i(\lambda^{(n)})&=&
\sum_j A_{ij}(\lambda^{(n)}) (\lambda_j^{(n+1)} -\lambda_j^{(n)})\ ,\\
b_i(\lambda^{(n)}) &=& 
\<S_i(\mbf{n})\>_u- \<\bar{S}_i(\mbf{n},\lambda^{(n)})\>_u\ ,\\
 A_{ij}(\lambda^{(n)})&=&
 \< \bar{S}_i\bar{S}_j -\overline{S_iS_j}\>_u\vert_{\lambda=\lambda^{(n)}}\ , 
\ea
where $\lambda^{(0)}_i=\tilde{\lambda}_i$ and
$ \lambda^{(n)}_i \rightarrow  \lambda_i$ as $n\rightarrow  \infty$.

The convergence depends on the choice of  the trial constants
$\tilde{\lambda}_i$. If iterations take many cycles, statistical errors
are likely to introduce instabilities in the solution. A similar problem
was encountered in \cite{ref}. The solution there was to find the iteration
limit through a linear mapping of the space of $\lambda_i$. Another method
to compute the coupling constants is to use the 
Schwinger-Dyson equations \cite{sd}.
In principle, the coupling constants can be compared with their ``exact''
values defined through (\ref{exacts}). The expectation value in the 
r.h.s. of (\ref{exacts}) can be expanded into a series over the spin
field around some specific spin field configuration. The expansion
coefficients can be computed my the Monte Carlo method in the Wilson
ensemble. For instance, the mass scale $\lambda_1$
can be obtained by taking the second derivative of  (\ref{exacts}) with 
respect to the spin field at the particular configuration $n_x^a=\delta^{3a}$
(as was suggested in the continuum case \cite{svs1}). This procedure
involves, however, computations of the determinant, which is very costly.  
Eq. (\ref{ym}) can be used to measure
the goodness of the approximation (\ref{action})--(\ref{s4}).
 
{\bf 5}. Having found the coupling constants, the renormalization group
flow for them has to be investigated to prove the stability of the 
effective action  in the limit of large
wave lengths. With this purpose, we use the representation (\ref{sphere})
to take into account the constraint $\mbf{n}_x^2 =1$. Let the matrix
$f_{xk}$ be a discrete Fourier transform associated with the lattice,
$\sum_x f^*_{k^\prime x}f_{xk}=\delta_{k^\prime k}$ and
$\sum_k f^*_{x^\prime k}f_{kx}=\delta_{x^\prime x}$.
The sum over $k$ implies the sum over all momentum vectors 
allowed by the lattice.
Given the ensemble of $\theta_x$ and $\phi_x$, the Fourier components
$\theta_k$ and $\phi_k$ can be computed. 

Next the spin field ensemble can be generated for all momenta bounded
from above by some scale $\Lambda_1$
\be
\theta_x(\Lambda_1) = \sum_{k\in K_1} f_{xk} \theta_{k}\ .
\label{ren}
\ee
Similarly for $\phi_x(\Lambda_1)$.  The sum in (\ref{ren}) is extended
over those vectors $k$ whose norm does not exceed the scale $\Lambda_1$.
This subset in the momentum space is denoted $K_1$. The ensemble
$\mbf{n}_x(\Lambda_1)$ can be used as the input for the inverse Monte-Carlo
procedure described in the previous section to compute new coupling
constants $\lambda_i(\Lambda_1)$. 

Repeating this procedure for successively smaller scales
$\Lambda_{k+1} < \Lambda_{k}$ we can obtain the sequence of the 
coupling constants $\lambda_i(\Lambda_k)$, $k=0,1,2,...$, where $k=0$
corresponds to the coupling constants computed with the original
spin field ensembles. By truncating the sum over momenta in (\ref{ren})
we generate a spin field ensemble in the infrared region (large wave
lengths). Hence the sequence  $\lambda_i(\Lambda_k)$ describes the behavior
of the coupling constants as functions of the 
scale $\Lambda$ that restricts allowed momenta from above in the
effective theory, i.e., as $\Lambda$ decreases, the infrared limit
is approached.  

The explicit removal of Fourier modes can be strongly affected 
by the breaking of rotational symmetry on coarse lattices or for 
larger $\Lambda$'s. So, the block spin decomposition \cite{sw,spin} might be 
a more attractive procedure  to study 
the renormalization group flow. The idea is to average spins over
elementary cells (blocks) of the original lattice. For instance, the angular
variables $\theta_x$ and $\phi_x$ are specified at elementary  
cubic cell vertices.
Consider a  new lattice with mesh $2a$ which is constructed as follows.
Let the point $y$ be the center of the elementary cell. The neighboring
sites are then $y\pm 2\mu$. So, each site $y$ of the new lattice
is inside an elementary cube $C_y$ of the original lattice, and the
cubes $C_y$ and $C_{y^\prime}$ do not have common vertices if $y\neq y^\prime$,
while $C_y$ and $C_{y^\prime}$ coincide if  $y=y^\prime$. 
Define
\be
\theta_y = 2^{-D-1}\sum_{x\in C_y}\theta_x   
\label{lav}
\ee
and similarly for $\phi_y$, where $D$ is the lattice dimension.
That is, $\theta_y$ is an average value of $\theta$ over all vertices
of one elementary cube. The spin
field $\mbf{n}_y$ is defined by (\ref{sphere}) where $x\rightarrow y$. 
The averaging (\ref{lav}) is also
equivalent to removing
short wave length components of the spin field. Doing this procedure
for a successively larger lattice spacing ($2a$, $4a$, etc.) 
and computing the coupling
constants on each step, we can again generate the renormalization
group flow $\lambda_i(\Lambda_k)$ (where $\Lambda_k \sim 2^{-k}/a$).  

The behavior of $\lambda_i(\Lambda)$ allows one to verify whether the 
effective action (\ref{1}) (possibly with extra terms) is
stable in the infrared region as was observed in \cite{cern} in the  
one-loop approximation. For instance, it is critical to observe
the right signs of $\lambda_{1,3}$ (cf. (\ref{1})) because 
$\lambda_1$ sets the mass scale for knot solitons, while $\lambda_3$
should reproduce the running gauge coupling constant $g$ in the continuum
limit. Relatively large (and growing)  
values of $ \lambda_{2,4}$ would mean instability
of knot solitons. It would also indicate that the approximation (\ref{1})
is not justified and the higher order spin interactions
are relevant for the spin field dynamics. In this case more higher order
terms have to be included into (\ref{action}) and the renormalization
group flow of the corresponding coupling constants is to be computed.   
      
It should also be noted that 
the lattice size becomes important in studies of $\lambda_i(\Lambda)$.
It is clear that  $\Lambda_{max}=\Lambda_0 \sim 1/a$ and 
$\Lambda_{min}\sim 1/L$ where $a$ is a lattice mesh and $L= Na$ with
$N$ being the number of lattice sites in one direction. Values 
of $\Lambda$ cannot be taken too close to $\Lambda_{min}$ because
the characteristic scaling behavior of the dimensional $\lambda_i$
in the continuum limit \cite{teper} must still be observable.

Let us summarize the essential steps of the algorithm. 
First, a Wilson ensemble
of the link variables is generated. Then
the spin field ensemble is computed. For every element of the ensemble
the functions $\bar{S}_i$ are computed for some trial coupling
constants. The integral over a spin field at a fixed site $x$ involved
in the definition of $\bar{S}_i$ has to be done numerically. 
Then the iterating algorithm is applied to compute the coupling
constants. Finally, the procedure is repeated for several ensembles
of the spin field which are obtained from the original ensemble 
by truncating (integrating out) short wave length components.
This gives the renormalization group flow of the coupling constants
which can tell us about the stability of the effective action in
the infrared limit. Using an appropriate scaling of $\lambda_1$ in 
the continuum limit, one can compute the mass scale of knot solitons   
supported by the effective action (\ref{1}), while the coefficients
at the additional terms would determine their stability.

As a final remark, it should be noted that the choice of the spin field
ensemble suggested above is associated with the maximal Abelian gauge.
It might be interesting to investigate  the effective 
action if different Abelian gauges are used to obtain  
the spin field ensemble. The purpose of such a study is to seek a numerical
evidence that the effective dynamics of Yang-Mills fields in the infrared
region is indeed governed by configurations that exhibit topological
defects when taken in any Abelian gauge \cite{pisa}. A generalization
of (\ref{nfield}) for the spin field ensemble can be defined as follows. 
Let $\Omega_x(u)$ be a gauge transformation such that $u_l^\Omega$
satisfies some Abelian gauge. The latter implies in particular that
there exists a field defined at each $x$, 
$\bar{\Phi}_x=\bar{\Phi}_x(u^\Omega)$ which is 
diagonal $[\tau_3, \bar{\Phi}_x]
=0$ and also $ \bar{\Phi}_x(u^\Omega) = \Omega_x^\dagger \Phi_x(n,u)\Omega_x$.
Therefore, a general equation for the spin field reads
\be
[\Phi_x(n,u), n_x] = 0\ ,
\label{nfieldg}
\ee
where $\Phi_x$ is any function of $n_x$ and $u_l$ which transforms according
to the adjoint representation of the gauge group. An example of the field
$\Phi_x(n,u)$ is given by the first two terms 
in Eq. (\ref{nfield}), and equation 
(\ref{nfield}) itself is clearly equivalent to (\ref{nfieldg}).
Link variables in two different Abelian gauges are related to each other
by a singular gauge transformation. Equation (\ref{nfieldg}) is covariant
under such transformations and, hence, represent the most general equation
for the spin field components.  
  
The simplest class of Abelian gauges is obtained by taking $\Phi_x$
to be independent of the spin field. In this case, the spin
field is identified as a solution of (\ref{nfieldg}): 
$n_x = \Phi_x(u)/|\Phi_x(u)|$ where
the vertical bars mean the norm with respect the trace scalar product
in the Lie algebra.
This is the case of the Polyakov Abelian gauge \cite{pag} and the so called
Laplacian Abelian gauge \cite{lag}. The spin field ensemble in the case
of the Polyakov Abelian gauge is probably the simplest one to generate.
In fact it is possible to average over all such choices of $\Phi$ in
the Yang-Mills path integral (dynamical Abelian gauge \cite{svs3}) 
by using a supersymmetric extension of the theory.  
 
The SU(N) generalization of (\ref{nfieldg})
is also straightforward. Take $N-1$ orthonormal and commutative spin
fields, $(n_x^a,n^b_x)=\delta^{ab}$ and $[n_x^a,n^b_x] =0$ where
$a,b =1,2, ..., N-1$. They should satisfy the equation $[\Phi_x(n,u),n_x^a]=0$
where $\Phi_x$ is some operator that transforms under the adjoint 
representation. 
In the simplest case when $\Phi_x$ depends only on the 
link variables, the solution for the spin fields is obtained
by orthogonalizing the commutative fields $\Phi_x^a\sim
d^{(a)}_{i_1i_2\cdots i_{a+1}}e^{i_1}\Phi_x^{i_2}\cdots \Phi_x^{i_{a+1}}$
where $e^i$ is an orthogonal basis in the Lie algebra $su(N)$, $\Phi_x
= e^i\Phi_x^i$ and   $d^{(a)}_{i_1i_2\cdots i_{a+1}}$ are the $N-1$ symmetric
irreducible invariant tensors of $su(N)$. Since for any Lie algebra element
$\Phi_x$ it always possible to construct $N-1$ linearly independent 
elements $\Phi_x^a$ (using the tensors $d^{(a)}$) that commute with
$\Phi_x$ and amongst each other,
the spin field can also be specified by  $[\Phi_x^b(n,u),n_x^a]=0$
for some choice of $\Phi_x^b(n,u)$.
A generalization of Eq. (\ref{nfield})
to the $SU(N)$ case is  simple. One has to replace $n_x$ by $n_x^a$
in the first two terms (this defines $\Phi_x^a(n,u)$) and 
$\xi_xn_x$ by $\sum_b \xi_x^{ab}n_x^b$ in the third one where
$\xi_x^{ab}= (\Phi_x^a,n_x^b)$.

{\bf Acknowledgments}. I am grateful to A. Wipf and his
colleagues for warm hospitality extended to me 
during my stay at the Institute for Theoretical Physics of the
University of Jena (Germany) and organizing a set of lectures
where some of the above ideas were first discussed. It is my
pleasure to thank the Alexander von Humboldt foundation 
for  financial support. I would like thank the referee for many
suggestions to improve the paper and useful references.   

{\bf Note added}. After submission of the paper for publication,
I have learned that the group of A. Wipf have done numerical simulations
on a $16^4$ lattice \cite{wipf}. The mass gap in the 
spin field spectrum and possible
global SO(3) symmetry  breaking (a preferable direction of the spin field)
have been reported so far. The existence of the mass gap around the lowest
glueball mass should certainly be expected because of the Abelian (and monopole)
dominance. The interesting question of the renormalization group flow
of the coupling constants is still to be studied.

\end{document}